\documentclass[aps,pra,twocolumn,showpacs]{revtex4}
\usepackage{graphicx}
\usepackage{bm}
\usepackage{amsmath}
\usepackage{amssymb}

\providecommand{\abs}[1]{\left\lvert#1\right\rvert}

\providecommand{\moy}[1]{\langle #1 \rangle}
\providecommand{\bra}[1]{\langle #1 \rvert}
\providecommand{\ket}[1]{\lvert #1 \rangle}

\begin{document}

\title{Quantum and classical separability of spin-orbit laser modes}
 
\author{L.~J.~Pereira, A.~Z.~Khoury and K.~Dechoum}

\affiliation{
Instituto de F\'\i sica, Universidade Federal Fluminense,
24210-346 Niter\'oi - RJ, Brasil.}

 
\begin{abstract}
In this work we investigate the quantum noise properties of polarization 
vortices in connection with an intensity based Clauser-Horne-Shimony-Holt 
inequality for their spin-orbit separability. 
We evaluate the inequality for different 
input quantum states and the corresponding intensity fluctuations. 
The roles played by coherence and photon number squeezing provide a 
suitable framework for characterizing pure state spin-orbit entanglement. 
Structural inseparability of the spin-orbit mode requires coherence, an 
issue concerning either classical or quantum descriptions. In both cases, 
it can be witnessed by violation of this intensity based CHSH inequality. 
However, in the quantum domain, entanglement requires both coherence and 
reduced photon number fluctuations. 

\end{abstract}
\vskip2pc 
 
\maketitle

\section{Introduction}
\label{introduction}

As a major resource for quantum information processing, entangled 
states have been intensively studied in a wide variety of quantum 
systems. From the mathematical point of view, when pure states are 
considered, they correspond to 
state vectors of a combined system that cannot be factorized as a 
tensor product of vectors belonging to the Hilbert spaces of 
the individual subsystems. In the general case, mixed 
entangled states cannot be written as a convex combination of 
projectors on factorized vectors. From the physical point of view, these 
states exhibit strong correlations when the individual systems are 
locally measured in different (non orthogonal) bases, as in the well 
known setting for the Clauser-Horn-Shimony-Holt (CHSH) inequality 
\cite{chsh}. For example, in photonic systems it has been tested 
with polarization entangled photon pairs \cite{aspect}. 
One can also encode a pair of qubits on the spin and orbital angular 
momentum of a single photon \cite{padgett1}. Combination of the 
two degrees of freedom leads to interesting applications on quantum 
information protocols including polarization controlled quantum 
imaging \cite{posinterf,imagepol}, algorithms \cite{deutschufmg}, 
gates \cite{marrucci1,marrucci2,marrucci3,cnotuff,sagnacuff}, 
cryptography \cite{cryptosteve,cryptouff,cryptolorenzo}, 
teleportation \cite{teleport1,teleport2,teleport3,teleport4}, 
and topological phases \cite{topoluff,topoluff2}. 
Of course, the CHSH inequality can be naturally applied to the 
spin-orbit degrees of freedom as well \cite{bellchines,belluff,bellkarimi}. 

Despite being frequently addressed in the quantum domain, 
tensor product structures are also present in the classical 
description of the electromagnetic field. We may quote the 
aforementioned example of the spin and orbital angular momentum 
carried by the optical field. While spin is related to the 
polarization, orbital angular momentum (OAM) has its origin 
in the wavefront structure \cite{padgettallen}. 
In a paraxial beam, these two 
degrees of freedom can be addressed independently and a 
combined mode basis can be built from a tensor product. 
Of course, one can also conceive non-separable spin-orbit 
modes and build a bridge for an 
interesting analogy with entangled states in quantum 
mechanics. Many concepts can be borrowed from quantum information theory, 
with a special role for the CHSH inequality as a non separability 
witness. This classical analogy of the CHSH inequality for 
optical degrees of freedom was suggested in Ref.\cite{spreew}. 
The spin-orbit version of the experiment was published in 
Ref.\cite{belluff} together with a theoretical description, 
including a brief quantum optical approach. Since then, other 
experimental \cite{saleh,jptorres} and theoretical 
\cite{simon,eberly,agarwal,dezela} works have addressed this issue. 
Spin-orbit entanglement has also been discussed in the quantum optical 
domain. The connection between classical inseparability and quantum 
entanglement was addressed in Ref.\cite{aiello1}, where the 
fundamental properties of cylindrically polarized modes were 
thoroughly investigated. Experimental tools for squeezing and entangling 
spin and orbital degrees of freedom were developed in 
Refs. \cite{aiello2,aiello3}. 

In this work, we discuss quantum optical aspects 
of the spin-orbit separability, with a special attention to the 
roles played by coherence and sub Poissonian photon number distribution 
\cite{zelaq1,zelaq2,zelaq3} in connection with the spin-orbit Bell 
measurements performed in Ref.\cite{belluff}. 
We will consider different input states, 
covering the main attributes directly related to the spin-orbit 
separability, especially coherence and quantum entanglement. The 
manuscript is organized as follows: in section \ref{polvortices} we review 
the basic description of polarization vortices as non-separable 
spin-orbit modes. These modes are then quantized in section \ref{qmodes} 
where different mode partitions are envisaged. The measurement scheme 
on which we base our theory is described in section \ref{bellmeasure} 
and the corresponding average intensities and quantum noise are derived 
in section \ref{qnoise} for different input states. Finally, we summarize 
our conclusions in section \ref{conclusion}.

\section{Polarization vortices as non-separable spin-orbit modes}
\label{polvortices}

In this section we discuss the notion of mode entanglement and 
its subtle distinction from quantum state entanglement. Both 
rely on the concept of non separability in a tensor product vector space. 
However, the vector structure of electromagnetic modes is not exclusive 
to quantum theory, it already occurs in classical electrodynamics. 
The polarization of a light beam is directly related to the vector nature 
of the electromagnetic field. At the same time, a vector space structure 
can be assigned to the spatial functions used to describe a paraxial 
beam. Therefore, the complete mode structure requires a combination of 
these two degrees of freedom in the form of a tensor product between 
the two vector spaces. For example, we can combine linear polarization 
vectors with Hermite-Gaussian spatial functions, which are discrete solutions  
of the paraxial wave equation in rectangular coordinates. 

We shall label the linear polarization directions with capital letters 
$H$ and $V$, associated with the unit vectors $\mathbf{\hat{e}}_H\equiv \mathbf{\hat{x}}$ and 
$\mathbf{\hat{e}}_V\equiv \mathbf{\hat{y}}\,$. The first-order Hermite-Gaussian modes oriented 
along the horizontal and vertical directions will be labeled with 
small case letters $h$ and $v$, respectively, associated with the 
spatial functions 
\begin{eqnarray}
\psi_h(x,y,z) &=& \mathcal{N}\,x\,\exp \left[ -\frac{(x^2+y^2)}{2w^2(z)}+i\phi(x,y,z) \right]\;,
\nonumber\\
\psi_v(x,y,z) &=& \mathcal{N}\,y\,\exp \left[ -\frac{(x^2+y^2)}{2w^2(z)}+i\phi(x,y,z) \right]\;,
\label{psihv}
\end{eqnarray}
where $\mathcal{N}$ is a normalization constant and $\phi(x,y,z)$ is the phase distribution. 
The most general first-order vector mode can be written as 
\begin{eqnarray}
\mathbf{\Psi}(\mathbf{r})&=& A_{Hh}\,\psi_h\,\mathbf{\hat{e}}_H + A_{Hv}\,\psi_v\,\mathbf{\hat{e}}_H + 
A_{Vh}\,\psi_h\,\mathbf{\hat{e}}_V 
\nonumber\\
&+& A_{Vv}\,\psi_v\,\mathbf{\hat{e}}_V\;, 
\label{Psi}
\end{eqnarray}
where $\sum_{\mu\nu} \abs{A_{\mu\nu}}^2=1\,$. 
We can borrow the definition of concurrence from quantum information theory to characterize 
the spin-orbit separability 
\begin{eqnarray}
\mathcal{C} = 2\,\left\| A_{Hh}\,A_{Vv}-A_{Hv}\,A_{Vh}\right\|\;, 
\label{concurrence}
\end{eqnarray}
so that $\mathcal{C}=0$ for product modes and $0<\mathcal{C}\leq 1$ for non factorisable ones. 

\begin{figure}[h!]
\includegraphics[scale=0.45]{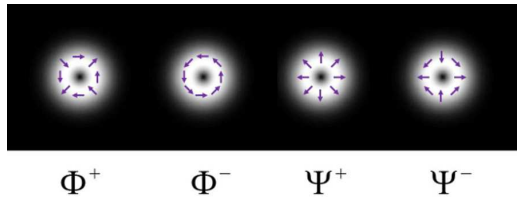}
\caption{\label{fig:modes} First order polarization vortices analogous to the four Bell states.}
\end{figure}

A first-order spin-orbit mode basis can also be built with maximally non-separable 
modes analogous to Bell states in quantum mechanics
\begin{eqnarray}
\mathbf{\Psi}_{\pm}(\mathbf{r})&=& \frac{\psi_h\,\mathbf{\hat{e}}_H \pm \psi_v\,\mathbf{\hat{e}}_V}{\sqrt{2}}\;, 
\nonumber\\ 
\mathbf{\Phi}_{\pm}(\mathbf{r})&=& \frac{\psi_h\,\mathbf{\hat{e}}_V \pm \psi_v\,\mathbf{\hat{e}}_H}{\sqrt{2}}\;.   
\label{bellbasis}
\end{eqnarray}
They correspond to non uniform polarizations on the wavefront. Two simple examples of maximally non 
separable modes are the well known radially ($\mathbf{\Psi}_{+}$) and azimuthally ($\mathbf{\Phi}_{-}$) 
polarized modes, characterized by the polarization patterns shown in Fig.(\ref{fig:modes}). Numerous methods for their 
production are available in the literature. They have attracted a considerable attention 
for their special focusing properties. It is easy to check from Eq.(\ref{concurrence}) that 
$\mathcal{C}=1$ for all elements of the basis given by Eq.(\ref{bellbasis}). 

\section{The quantized spin-orbit modes}
\label{qmodes}

Let us consider the Heisenberg operator describing the positive frequency 
component of the electric field associated with a first-order paraxial beam 
propagating along $z\,$. In the separable mode basis 
$\{\psi_\nu(\mathbf{r})\,\mathbf{\hat{e}}_\mu\}\,$, with $\mu=H,V$ and 
$\nu=h,v\,$, it reads 
\begin{eqnarray}
\mathbf{A}^{(+)}(\mathbf{r},t)=e^{i\left(kz-\omega t\right)}\,
\sum_{\mu,\nu} a_{\mu\nu}\,\psi_\nu(\mathbf{r})\,\mathbf{\hat{e}}_\mu \;,
\label{Ebell}
\end{eqnarray}
where $a_{\mu\nu}$ is the operator that annihilates a photon on mode 
$\psi_\nu(\mathbf{r})\,\mathbf{\hat{e}}_\mu\,$. 
As usual, we can also define new annihilation operators when a different 
mode basis is used to decompose the field operator. For the Bell modes 
defined in section \ref{polvortices}, we obtain the following transformation 
equations for annihilation operators
\begin{eqnarray}
a_{\Psi_{\pm}} &=& \frac{a_{Hh} \pm a_{Vv}}{\sqrt{2}}\;, 
\nonumber\\ 
a_{\Phi_{\pm}} &=& \frac{a_{Hv} \pm a_{Vh}}{\sqrt{2}}\;, 
\label{annihibell}
\end{eqnarray}
and the corresponding conjugate transformations for the creation operators. 
The electric field operator can then be written as 
\begin{eqnarray}
\mathbf{A}^{(+)}(\mathbf{r},t) &=& e^{i\left(kz-\omega t\right)}\,\sum_{\pm}
a_{\Psi_{\pm}}\,\mathbf{\Psi}_{\pm} + a_{\Phi_{\pm}}\,\mathbf{\Phi}_{\pm}\;. 
\label{Ebell}
\end{eqnarray}

Now it is interesting to investigate the transformations for different kinds 
of states. A Fock basis of the Hilbert space can be built for either mode 
decompositions by acting on the vacuum state with the corresponding creation operators.   
Let us consider, for example, an $N-$photon Fock state on mode $\mathbf{\Psi_+}\,$, 
with all other modes empty. For simplicity, we shall omit the empty modes in the state 
vector, unless they are directly involved in the transformations that we are 
interested, however we must keep in mind that we will be always dealing with a multimode 
Hilbert space. Thus, for the mode in question we have 
\begin{eqnarray}
\ket{N}_{\Psi_+}\,\ket{0}_{\Psi_-} = \frac{\left(a_{\Psi_+}^{\dagger}\right)^N}{\sqrt{N!}}\,\,\ket{vac}\;.
\label{fockpsi+}
\end{eqnarray}
If we use the conjugate transformations (\ref{annihibell}) and write the creation operator 
for $\mathbf{\Psi}_+$ in terms of those for $Hh$ and $Vv$, we obtain 
\begin{eqnarray}
\ket{N}_{\Psi_+}\,\ket{0}_{\Psi_-} = 
\sum_{n=0}^{N}\frac{\sqrt{N!}\left(a_{Hh}^{\dagger}\right)^n\left(a_{Vv}^{\dagger}\right)^{N-n}}
{2^{N/2}\,\,n!\,(N-n)!}
\,\,\,\,\ket{vac}\;.
\nonumber\\
\label{fockpsi+2}
\end{eqnarray}
From the analogous definitions of the Fock states on modes $Hh$ and $Vv\,$, we readily identify them 
in the right hand side of Eq.(\ref{fockpsi+2}), so that  
\begin{eqnarray}
\ket{N}_{\Psi_+}\,\ket{0}_{\Psi_-} = 
\sum_{n=0}^{N}\,\sqrt{\frac{N!\,/2^N}{n!\,(N-n)!}}\,\,\ket{n}_{Hh}\,\ket{N-n}_{Vv}\;.
\nonumber\\
\label{fockpsi+3}
\end{eqnarray}
For example, a single photon Fock state on mode $\mathbf{\Psi}_+\,$ can be written 
\begin{eqnarray}
\ket{1}_{\Psi_+}\,\ket{0}_{\Psi_-} = \frac{\ket{1}_{Hh}\,\ket{0}_{Vv}+\ket{0}_{Hh}\,
\ket{1}_{Vv}}{\sqrt{2}}\;.
\label{fockpsi+1photon}
\end{eqnarray}
Note that the left hand side of Eqs. (\ref{fockpsi+3}) and (\ref{fockpsi+1photon}) 
has a product state while 
the right hand side has an entangled one. Of course, entanglement is independent of 
the basis chosen for the Hilbert space, however these equations are \textit{not} 
a change of basis, they correspond to different mode partitions of the electromagnetic 
field. Therefore, the mode basis must not be confused with the quantum state basis. 
In this sense, we prefer to employ the term \textit{mode partition} in order to avoid 
this kind of confusion. 

Entanglement does depend on the way the subsystems are defined and, although 
this partition is usually evident for material systems, it is not so clear for the 
electromagnetic field since one often has more than one mode decomposition adapted 
to the same boundary conditions. Therefore, we must be careful when comparing entanglement 
in the two cases. In this respect, pure coherent states play a very special role because 
they are \textit{always} a product state in whatever mode decomposition. 
Indeed, let 
\begin{eqnarray}
D_{\Psi_+}(u)=\exp\left(u\,a_{\,\Psi_+}^\dagger - u^*\,a_{\,\Psi_+}\right)
\label{DPsi+}
\end{eqnarray}
be the displacement operator that produces the coherent state 
$\ket{u}_{\Psi_+}\,\ket{0}_{\Psi_-}$ 
when acting on the vacuum state. Since $a_{Hh}$ and $a_{Vv}$ commute, 
it can be factorized as a product between the corresponding displacement 
operators of modes $Hh$ and $Vv\,$. From Eqs.(\ref{annihibell}) one 
easily obtains 
\begin{eqnarray}
D_{\Psi_+}(u)=D_{Hh}(u/\sqrt{2})\,D_{Vv}(u/\sqrt{2})\;, 
\label{DPsi+}
\end{eqnarray}
so that the coherent states in the two mode decompositions are related by 
\begin{eqnarray}
\ket{u}_{\Psi_+}\,\ket{0}_{\Psi_-} = \ket{u/\sqrt{2}}_{Hh}\,\ket{u/\sqrt{2}}_{Vv}\;, 
\label{coherentpsi+}
\end{eqnarray}
which is a product state in any case. 
From this discussion we develop some intuition about the role played by photon number 
noise in the interplay between mode separability and quantum entanglement. We next 
discuss this relationship more closely in the context of the experiment done in 
Ref.\cite{belluff}.

\section{Spin-orbit Bell measurements}
\label{bellmeasure}

\begin{figure}[h!]
\includegraphics[scale=0.5]{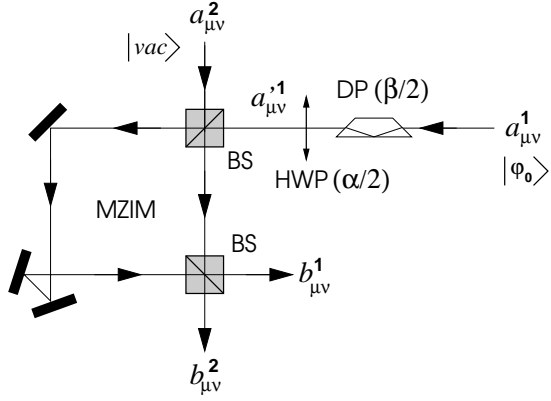}
\caption{\label{fig:interferometer} Scheme for the spin-orbit Bell measurement.}
\end{figure}

Let us consider the sketch shown in Fig.(\ref{fig:interferometer}) for a spin-orbit Bell measurement 
as the one done in \cite{belluff}. An input mode initially prepared in some quantum 
state $\ket{\varphi_0}$ enters a Mach-Zehnder interferometer with an extra mirror in one arm. 
This type of interferometer will be designated by the acronym MZIM \cite{sasada}. 
When the two arms are equilibrated, the interferometer sorts even and odd modes 
at different output ports, the components $Hh$ and $Vv$ exit from the even port while 
$Hv$ and $Vh$ exit from the odd port. 
The input beams are described by the electric field operators 
\begin{eqnarray}
\mathbf{A}_{j}^{(+)}(\mathbf{r},t)=e^{i\left(\mathbf{k}_j\cdot\mathbf{r}-\omega t\right)}\,
\sum_{\mu,\nu} a^{j}_{\mu\nu}\,\psi_\nu(\mathbf{r})\,\mathbf{\hat{e}}_\mu \;,
\end{eqnarray}
where $\mu=H,V$ and $\nu=h,v$ refer to polarization and transverse mode orientation respectively, $j=1,2$ refers to the input port, 
and $a^j_{\mu\nu}$ is the corresponding annihilation operator.  
The input field coming through port 1 will be transmitted through unitary transformation elements acting on the 
transverse mode and polarization degrees of freedom. These elements are used to set the bases of the Bell measurement. They 
can be a Dove prism (DP) for transverse mode and a half wave plate (HWP) for polarization transformation. When the HWP is 
oriented at an angle $\alpha/2$ and the DP at $\beta/2\,$, the annihilation operators are transformed according to 
\begin{eqnarray}
a^{{\prime\,1}}_{\mu\nu}(\alpha,\beta)=\sum_{\epsilon\delta} U_{\mu\nu}^{\epsilon\delta}\,a^{1}_{\epsilon\delta}\;,
\end{eqnarray}
where $U$ is the unitary transformation describing the combined action of the two elements. The HWP and the DP 
both act in a similar manner, they produce a reflection operation along their orientation axis. Therefore, their 
combined action can be written as the tensor product $U=T(\alpha)\otimes T(\beta)\,$, where 
\begin{eqnarray}
T(\theta) = \left(
\begin{matrix}
\cos\theta & \sin\theta\\
\sin\theta & -\cos\theta
\end{matrix}
\right)\;.
\label{uw}
\end{eqnarray}
In matrix notation we have
\begin{equation}
\left(
\begin{matrix}
a_{Hh}^{\prime\,1}\\
a_{Hv}^{\prime\,1}\\
a_{Vh}^{\prime\,1}\\
a_{Vv}^{\prime\,1}
\end{matrix}
\right)
 = \left(
\begin{matrix}
\cos\alpha & \sin\alpha\\
\sin\alpha & -\cos\alpha
\end{matrix}
\right)
\otimes
\left(
\begin{matrix}
\cos\beta & \sin\beta\\
\sin\beta & -\cos\beta
\end{matrix}
\right)\;
\left(
\begin{matrix}
a_{Hh}^1\\
a_{Hv}^1\\
a_{Vh}^1\\
a_{Vv}^1
\end{matrix}
\right)\;.
\label{atoc}
\end{equation}

The same decomposition also applies to the output field operators 
\begin{eqnarray}
\mathbf{B}_{j}^{(+)}(\mathbf{r},t)=e^{i\left(\mathbf{k}_j\cdot\mathbf{r}-\omega t\right)}\,
\sum_{\mu,\nu} b^j_{\mu\nu}\,\psi_\nu(\mathbf{r})\,\mathbf{\hat{e}}_\mu \;, 
\end{eqnarray}
where $b^j_{\mu\nu}$ is the annihilation operator corresponding to mode $\mu\nu$ at output $j\,$. 

One of the output ports of the interferometer will combine the odd modes from the vacuum 
input with the even modes from the excited input and the opposite combination is 
performed in the other output port. The quantized mode amplitudes $b_{\mu\nu}$ are 
then determined by the input-output relations for the interferometer
\begin{eqnarray}
b^1_{Hh}&=&a^{\prime\,1}_{Hh}\,, \;\;\;\;\;\;\;\;\;\;  b^2_{Hh}=a^2_{Hh}\,, \nonumber\\
b^1_{Hv}&=&a^2_{Hv}\,, \;\;\;\;\;\;\;\;\;\;  b^2_{Hv}=a^{\prime\,1}_{Hv}\,, \nonumber\\
b^1_{Vh}&=&a^2_{Vh}\,, \;\;\;\;\;\;\;\;\;\;  b^2_{Vh}=a^{\prime\,1}_{Vh}\,, \nonumber\\
b^1_{Vv}&=&a^{\prime\,1}_{Vv}\,, \;\;\;\;\;\;\;\;\;\;  b^2_{Vv}=a^2_{Vv}\;. 
\end{eqnarray}

We now define the intensity operators integrated over the detectors area 
\begin{eqnarray}
\hat{I}_j=\int_{D_j} \mathbf{B}_{j}^{(-)}\cdot\mathbf{B}_{j}^{(+)}\,\,d^2\mathbf{r}\;. 
\end{eqnarray}
From the input-output relations of the measurement apparatus (transformation elements 
plus interferometer) one finds
\begin{eqnarray}
\hat{I}_1&=&\hat{I}^1_{Hh}+\hat{I}^1_{Vv}+\hat{I}^1_{Hv}+\hat{I}^1_{Vh}\;,
\nonumber\\
\hat{I}_2&=&\hat{I}^2_{Hv}+\hat{I}^2_{Vh}+\hat{I}^2_{Hh}+\hat{I}^2_{Vv}\;,
\end{eqnarray}
where $\hat{I}^j_{\mu\nu}=b_{\mu\nu}^{j\,\dagger}\,b_{\mu\nu}^j\,$. 
Note that modes $Hv$ and $Vh$ on output port 1 are not excited, as well as 
modes $Hh$ and $Vv$ on output port 2. 
Therefore, they do not contribute to intensity measurements 
and the relevant operators at the output of the 
interferometer are  
\begin{eqnarray}
I_{tot} &=& \hat{I}_1 + \hat{I}_2\;,
\nonumber\\
M(\alpha,\beta) &=& \hat{I}_1 - \hat{I}_2\;.
\end{eqnarray}
The total input-output intensity does not depend on 
the measurement settings $(\alpha,\beta)\,$.

In analogy with the parameter used in the Clauser-Horn-Shimony-Holt (CHSH) inequality, 
we define 
\begin{eqnarray}
S=
\frac{\moy{\,M(\alpha,\beta) + M(\alpha,\beta^\prime)
- M(\alpha^\prime,\beta) + M(\alpha^\prime,\beta^\prime)\,}}
{\moy{I_{tot}}}\;, 
\nonumber\\
\end{eqnarray}
where the averages are calculated with the initial quantum state $\ket{\varphi_0}\,$. 
For single photon states, the normalized detected intensities correspond to 
detection probabilities and one finds the usual scenario for Bell measurement where 
$2<S<2\sqrt{2}$ for single photon spin-orbit entangled states. As discussed in Ref.\cite{belluff}, 
this violation can be achieved with a polarization vortice coherent state, which is a 
product state in the $Hh$ and $Vv$ decomposition, as shown in Eq.(\ref{coherentpsi+}). 
Therefore, in order to evidence quantum behavior, it will be necessary to monitor 
the intensity difference noise $\Delta M^2(\alpha,\beta)=\moy{M^2(\alpha,\beta)}-\moy{M(\alpha,\beta)}^2$ 
and compare it with the corresponding shot noise limit. 
This figure of merit will be used in the next section 
to set up the comparison between different input states.

\section{Intensity averages and quantum noise}
\label{qnoise}

Let us assume that only modes $a^1_{Hh}$ and $a^1_{Vv}$ are initially occupied with 
some quantum state $\ket{\varphi_0}\,$, while all other modes are empty. 
We will proceed as follows, first all output intensity operators are 
calculated in terms of the input modes (before HWP and DP)  
using the input-output relations outlined above. Then the mean values 
giving the average intensity and the corresponding quantum fluctuation 
are calculated with the specified initial state. We will discuss different 
input states, their similarities and differences.

\subsection{Entangled Fock states}
\label{qnoisefock}

Let us first consider the single photon entangled state given by Eq.(\ref{fockpsi+1photon}). 
In this case, we have 
\begin{eqnarray}
\frac{\moy{M(\alpha,\beta)}}{\moy{I_{tot}}}=\cos\,[2(\beta-\alpha)]\;, 
\label{fockM} 
\end{eqnarray}
where $\moy{I_{tot}}=1\,$. 
For the measurement settings $\alpha=\pi/8\,$, $\alpha^\prime=3\pi/8\,$, $\beta=0\,$, $\beta^\prime=\pi/4\,$, 
we obtain the limiting value $S=2\sqrt{2}$ expected for a maximally spin-orbit entangled state. 
We also evaluate the corresponding intensity difference fluctuations for any setting $(\alpha,\beta)$ 
\begin{eqnarray}
\frac{\Delta M^2(\alpha,\beta)}{\moy{I_{tot}}}=\sin^2\,[2(\beta-\alpha)]\;.
\label{fockDeltaM} 
\end{eqnarray}
These results can be easily generalized for the arbitrary $N-$photon entangled state $\ket{N}_{\Psi_+}\,\ket{0}_{\Psi_-}$ 
given by Eq.(\ref{fockpsi+3}) for which $\moy{I_{tot}}=N\,$. 
For the Bell measurement settings we obtain $50\%$ intensity squeezing. 
When $\beta-\alpha=m\,\pi/2\,(m\in \mathbb{Z})\,$, all photons exit through the same output port and 
the intensity difference is perfectly squeezed. 

Note that inter-mode coherence is a key ingredient for violation of the CHSH inequality, since the 
summations on the right hand side of Eqs. (\ref{fockpsi+3}) and (\ref{fockpsi+1photon}) involve 
a coherent superposition with well defined relative phases. When those are randomized, one gets 
statistical mixtures of Fock states which no longer violate the inequality, as we will see next.

\subsection{Mixed Fock states}
\label{qnoisemixfock}

It will be interesting to evaluate the spin-orbit CHSH inequality for a mixed Fock state 
exhibiting strong photon number correlations but no entanglement. This state can be 
represented by the density matrix
\begin{eqnarray}
\rho_{N} = 
\sum_{n=0}^{N}\,\frac{N!\,/2^N}{n!\,(N-n)!}\,\,\ket{n}\bra{n}_{Hh}\otimes\ket{N-n}\,\bra{N-n}_{Vv}\;,
\nonumber\\
\label{fockmixed}
\end{eqnarray}
with a well defined total photon number $N$, randomly distributed between modes $Hh$ and $Vv\,$. 
For this state we obtain 
\begin{eqnarray}
\frac{\moy{M(\alpha,\beta)}}{\moy{I_{tot}}} = \cos\,2\alpha\,\cos\,2\beta\;, 
\nonumber\\  
\label{fockmixedM}
\end{eqnarray}
what results in $S=\sqrt{2}$ for the measurement settings. 

The intensity difference noise is
\begin{eqnarray}
\frac{\Delta M^2(\alpha,\beta)}{\moy{I_{tot}}} &=& \sin^22\alpha + \sin^22\beta
\nonumber\\ 
&+&\left(\frac{\moy{I_{tot}}-3}{2}\right)\,\sin^22\alpha\,\sin^22\beta \;, 
\label{fockmixedDeltaM}
\end{eqnarray}
which can exhibit perfect squeezing for specific settings $(\alpha,\beta)$, but in 
general scales with the photon number. For example, if we set $\alpha=\beta=\pi/4\,$, 
expression (\ref{fockmixedDeltaM}) gives $(N+1)/2$ while (\ref{fockDeltaM}) predicts 
perfect squeezing.  

\subsection{Photon number Werner states}
\label{qnoisewerner}

In order to capture the role played by entanglement in the quantum noise properties 
of Fock states, let us consider a partially entangled state of the kind
\begin{eqnarray}
\rho_N(p) = p\,\ket{N}\bra{N}_{\Psi_+}\otimes\ket{0}\bra{0}_{\Psi_-} + (1-p)\,\rho_N\;,
\label{wernerfock}
\end{eqnarray}
where $\rho_N$ is given by Eq.(\ref{fockmixed}). It is analogous to the 
Werner states frequently used in quantum information science to discuss correlations 
present in partially entangled states \cite{werner}. The average intensity difference for 
setting $(\alpha,\beta)$ is 
\begin{eqnarray}
\frac{\moy{M(\alpha,\beta)}}{\moy{I_{tot}}} = p\,\cos\,\left[2(\beta-\alpha)\right] + 
(1-p)\,\cos\,2\alpha\,\cos\,2\beta\;, 
\nonumber\\  
\label{wernerfockM}
\end{eqnarray}
which gives $S=(1+p)\sqrt{2}\,$, corresponding to a violation threshold at $p=\sqrt{2}-1\,$.  

The intensity noise can be readily calculated as a combination of the results given by 
Eqs.(\ref{fockDeltaM}) and (\ref{fockmixedDeltaM}). In fact, one can easily verify the 
relation 
\begin{eqnarray}
\moy{\Delta M^2}_W &=& p\,\moy{\Delta M^2}_{pure} + (1-p)\,\moy{\Delta M^2}_{mix}
\nonumber\\
&+& p\,(1-p)\,\left[\,\moy{M}_{pure}-\moy{M}_{mix}\right]^{\,\,2}\,, 
\label{wernerDM} 
\end{eqnarray}
where $\moy{\Delta M^2}_{pure}$ is given by Eq.(\ref{fockDeltaM}), which scales as 
$\moy{I_{tot}}$ while $\moy{\Delta M^2}_{mix}$ is given by Eq.(\ref{fockmixedDeltaM}), 
which scales as $\moy{I_{tot}}^2\,$. Note that both $\moy{M}_{pure}$ and 
$\moy{M}_{mix}$ are proportional to $\moy{I_{tot}}\,$, so that 
$\moy{\Delta M^2}_{W}/\moy{I_{tot}}$ scales as $\moy{I_{tot}}\,$.

\subsection{Pure coherent state}
\label{qnoisepurecoherent}

For a two-mode coherent state $\ket{u}_{\Psi_+}\,\ket{0}_{\Psi_-}$ given by Eq.(\ref{coherentpsi+}) 
we obtain 
\begin{eqnarray}
\frac{\moy{M(\alpha,\beta)}}{\moy{I_{tot}}}=\cos\,[2(\beta-\alpha)]\;,  
\label{purecoherentM}
\end{eqnarray}
with $\moy{I_{tot}}=\abs{u}^2\,$, giving the same limiting value 
$S=2\sqrt{2}$ as for the entangled Fock state, however, the input state is now factorized. 
In order to distinguish the two cases, we calculate the intensity difference noise to find  
\begin{eqnarray}
\frac{\Delta M^2(\alpha,\beta)}{\moy{I_{tot}}}=1\;, 
\label{purecoherentDM}
\end{eqnarray}
so that shot noise is expected for any $(\alpha,\beta)$ settings, which is radically 
distinguished from the corresponding result for entangled Fock states.

\subsection{Mixed coherent state}
\label{qnoisemixcoherent}

\begin{figure}[h!]
\includegraphics[scale=0.45]{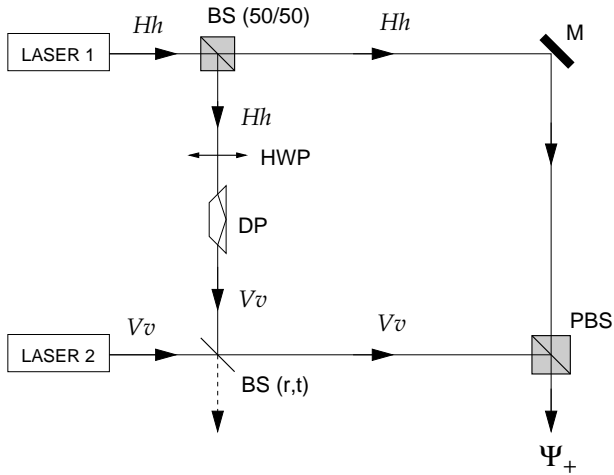}
\caption{\label{fig:interferometer2} Proposed scheme to produce an adjustable mixed coherent state.}
\end{figure}

We now investigate the role played by coherence in the violation of the spin-orbit 
CHSH inequality. Suppose a laser beam is initially prepared in a coherent state 
$\ket{u}_{Hh}\,\ket{0}_{Vv}\,$. This beam is then split in two equal parts, 
one of them is transformed to mode $Vv$ and sent to a beam splitter 
where it gets \textit{contaminated} with an independent 
laser also prepared in $Vv\,$. Finally, the mixed $Vv$ mode is recombined with 
$Hh$ on a polarizing beam splitter (PBS) as sketched in Fig.(\ref{fig:interferometer2}). 
Since the two lasers have random relative phase, the quantum state describing the 
final beam can be written as the following statistical mixture
\begin{eqnarray}
\rho_{0} &=& \ket{u}\bra{u}_{Hh}\otimes 
\int\,\frac{d\theta}{2\pi}\,\,\ket{u^\prime(\theta)}\bra{u^\prime(\theta)}_{Vv}\;, 
\label{partcoherent}
\end{eqnarray}
where $u^\prime(\theta) = u\,(r+t\,e^{i\theta})\,$, with $r$ and $t$ being the reflection 
and transmission coefficients of the contaminating coupler. 
Let $r=\sqrt{R}\,e^{i\phi}\,$, where $R$ is the 
intensity reflectivity and $\phi$ is the phase acquired on reflection, 
then we obtain 
\begin{eqnarray}
\frac{\moy{M(\alpha,\beta)}}{\moy{I_{tot}}} = \cos\,2\alpha\,\cos\,2\beta + 
\sqrt{R}\,\cos\phi\,\sin 2\alpha\sin 2\beta\;,
\nonumber\\  
\end{eqnarray}
where $\moy{I_{tot}}=2\abs{u}^2\,$, giving $S=(1+\sqrt{R}\,\cos\phi)\sqrt{2}$ 
for the Bell measurement settings. A violation threshold is predicted at 
$\sqrt{R}\,\cos\phi=\sqrt{2}-1\,$. 
For $R=1$ and $\phi=0$ one recovers the coherent state result given by 
Eq.(\ref{purecoherentM}). 

For a completely incoherent superposition ($R=0$) we obtain 
\begin{eqnarray}
\frac{\moy{M(\alpha,\beta)}}{\moy{I_{tot}}}=\cos\,2\alpha\,\cos\,2\beta\;,  
\end{eqnarray}
giving $S=\sqrt{2}$ for the Bell measurement settings. This result 
suggests that coherence is an essential ingredient for violation 
of the inequality for the intensity average. Meanwhile, the intensity 
noise for a setting $(\alpha,\beta)$ is 
\begin{eqnarray}
\frac{\Delta M^2(\alpha,\beta)}{\moy{I_{tot}}} &=&
1+\frac{\moy{I_{tot}}}{2} \sin^22\alpha\,\sin^22\beta\;, 
\end{eqnarray}
which scales as $\moy{I_{tot}}$ and is above shot noise for most settings.

\subsection{Two-mode squeezed vacuum}
\label{qnoisesqueezed}

It is also interesting to inspect the results obtained for the two-mode 
squeezed vacuum state, which presents quadrature entanglement.  
It can be constructed from the action of the two-mode squeezing 
operator on the vacuum state as given by 
\begin{eqnarray}
\ket{\varphi_{sq}}=\exp\left(\frac{\zeta^*\,a_{Hh}\,a_{Vv}-\zeta\,a^\dagger_{Hh}\,a^\dagger_{Vv}}{2}\right)
\ket{vac}\;, 
\label{2squeezed} 
\end{eqnarray}
where $\zeta$ is the squeezing parameter and the total intensity is $\moy{I_{tot}}=2\,\sinh^2(\,\abs{\zeta}/\,2)\,$. 
In this case we find 
\begin{eqnarray}
\frac{\moy{M(\alpha,\beta)}}{\moy{I_{tot}}}=\cos\,2\alpha\,\cos\,2\beta\;,  
\end{eqnarray}
giving $S=\sqrt{2}$ for the measurement settings, the same result as for the mixed coherent state.
For the intensity noise we get 
\begin{eqnarray}
\frac{\Delta M^2(\alpha,\beta)}{\moy{I_{tot}}}=1+
\left(\moy{I_{tot}}+1\right)\left(\frac{1+\cos 4\alpha\,\cos 4\beta}{2}\right)\;, 
\nonumber\\
\label{delta2squeezed}
\end{eqnarray}
which also scales as $\moy{I_{tot}}$ for most settings. Therefore, despite its quadrature 
entanglement, the quantum properties of squeezed vacuum do not show up in this 
kind of spin-orbit Bell measurement.

\section{conclusion}
\label{conclusion}

In summary, we have discussed spin-orbit Bell measurements in paraxial polarization 
vortices from a quantum optical perspective. The vortices are treated as non-separable 
modes and their quantum descriptions in different mode decompositions are compared. 
The mode separability is evaluated with an inequality for the average intensities 
analogous to the CHSH inequality. It is violated for a single-photon 
spin-orbit entangled state, as expected, but also by a factorized 
coherent state. These two limiting cases were experimentally investigated in Refs.
\cite{bellchines,bellkarimi,belluff}. Therefore, a natural question arises about 
the interplay between entanglement and coherence in the violation. In this article 
we compared different input states in order to develop some intuition. 

Our results indicate that the roles played by coherence and photon number 
fluctuations can be summarized by the following examples: 

\begin{itemize}
\item 
Without coherence we expect no violation of the intensity based CHSH 
inequality, as illustrated by the examples with mixed coherent or Fock 
states (sections \ref{qnoisemixfock} and \ref{qnoisemixcoherent}).  

\item
With coherence and Poissonian distribution for the total photon number, 
we expect violation but no entanglement, as illustrated by the example 
with product coherent states  (section \ref{qnoisepurecoherent}). We 
remark that pure coherent states are factorized in any mode decomposition 
but exhibit maximal violation.

\item
With coherence and reduced (below shot-noise) photon number fluctuations, 
we expect violation and entanglement, as illustrated by the entangled 
Fock states (section \ref{qnoisefock}). 
\end{itemize}
It is important to stress that in all cases the relevant coherence being 
considered is the one characterized by inter-mode relative phase, not 
the absolute one. 
We also present examples that interpolate between pure and mixed states. 
Finally, section \ref{qnoisesqueezed} illustrates that quadrature 
entanglement does not imply in violation of the intensity based CHSH 
inequality.

Experimentally, structural inseparability and entanglement can be 
simultaneously certified by measuring intensity averages and 
the corresponding fluctuations. Violation of the CHSH inequality 
with the averages certify structural inseparability while intensity 
difference squeezing evidences entanglement.  
Coherent states should exhibit shot noise level for any measurement 
settings while photon number squeezed states should present 50\% 
squeezing for the settings used in the inequality. 
These results can be experimentally tested with polarization vortices produced 
on intensity squeezed sources like optical parametric oscillators \cite{fabre,hyper} 
or pump-noise-supressed diode lasers \cite{yamamoto}. 

\section*{Acknowledgments}
Funding was provided by Coordena\c c\~{a}o de Aperfei\c coamento de 
Pessoal de N\'\i vel Superior (CAPES), Funda\c c\~{a}o de Amparo \`{a} 
Pesquisa do Estado do Rio de Janeiro (FAPERJ-BR), and Instituto Nacional 
de Ci\^encia e Tecnologia de Informa\c c\~ao Qu\^antica (INCT-CNPq).

\end{document}